\documentclass[english]{lni}
\pdfoutput=1

\usepackage[hyphens]{url}
\usepackage[utf8]{inputenc} 
\usepackage{graphicx}
\usepackage{amssymb}
\usepackage{pifont}
\usepackage{fancyhdr}
\usepackage{breakurl}
\usepackage{hyperref}
\usepackage{listings} 
\usepackage{changepage} 
\usepackage[figurename=Fig., tablename=Tab., small]{caption}[2008/04/01]
\usepackage{xcolor}
\definecolor{bluekeywords}{rgb}{0,0,1}
\definecolor{greencomments}{rgb}{0,0.5,0}
\definecolor{redstrings}{rgb}{0.64,0.08,0.08}
\definecolor{xmlcomments}{rgb}{0.5,0.5,0.5}
\definecolor{types}{rgb}{0.17,0.57,0.68}

\newcommand{\xmark}{\ding{55}}

\renewcommand{\headrulewidth}{0.4pt} 
\author{Jakob Rieck\footnote{University of Hamburg, FB Informatik, Security in Distributed Systems Group, Vogt-Kölln-Straße 30, 22527 Hamburg, 2rieck@informatik.uni-hamburg.de}}
\title{Attacks on Fitness Trackers Revisited:\\ A Case-Study of Unfit Firmware Security}
\begin{document}
\maketitle
\setcounter{footnote}{1}

\begin{abstract}
Fitness trackers -- wearables that continuously record a wearer's step count and related activity data -- are quickly gaining in popularity. Apart from being useful for individuals seeking a more healthy lifestyle, their data is also being used in court and by insurance companies to adjust premiums. For these use cases, it is essential to ensure authenticity and integrity of data. Here we demonstrate a flaw in the way firmware for Withings' \emph{Activité} is verified, allowing an adversary to compromise the tracker itself. This type of attack has so far not been applied to fitness trackers. Vendors have started mitigating previous attacks, which manipulated data by interfering with wireless channels, or by physically moving the tracker to fool sensors. Hardware similarities amongst different trackers suggest findings can be transferred to other tracker as well.
\end{abstract}
\begin{keywords}
Fitness Tracker, IoT, Wearable Security, Withings, Firmware Modification, Reverse Engineering
\end{keywords}

\section{Introduction}
\label{introduction}

Fitness trackers -- wearable devices that capture various statistics about their wearer and try to motivate physical activity -- have become increasingly popular in recent times. According to Canalys, a market research firm, wearable fitness bands grew 684\% worldwide in the first half of 2014 compared to the first half of 2013 \cite{canalys}. Estimations for future growth vary wildly, indicating the sector is not yet well understood and possibly grows much faster than older, more traditional sectors \cite{Wei14}.

Fitness trackers are typically wrist-worn devices that collect and send data such as the wearer's step count via a user's device to the companies' server. Existing research indicates that generated data, despite being noisy, can be used to answer intimate questions, such as if a person recently quit smoking \cite{Ka14} or whether two humans work together \cite{tsubouchi2013working}.

Insurance companies have started to offer benefits for active customers who choose to share their fitness data with them \cite{Be15}. Data generated by fitness trackers is used in court rooms and can decide the outcome of a trial \cite{Hi15, Ol14}.
Both use-cases raise questions about how the data's \emph{integrity} and \emph{authenticity} can be ensured. If a vendor fails to address these issues, insurance fraud starts to become a problem and fabricated evidence ends up deciding cases in court. Existing technical attacks try to manipulate data in transit and can be effectively mitigated by affected vendors. Firmware attacks, seemingly overlooked in this context, pose a new challenge, as they violate trust placed in hardware.

\clearpage
\pagestyle{fancy}
\fancyhead{} 
\fancyhead[RO]{\small Attacks on Fitness Trackers Revisited \hspace{5pt} \thepage \hspace{0.05cm}}
\fancyhead[LE]{\hspace{0.05cm}\small \thepage \hspace{5pt} Jakob Rieck}
\fancyfoot{} 
\renewcommand{\headrulewidth}{0.4pt} 

This paper contributes to the field as follows: In the first part, a general architectural model of fitness trackers is presented and existing attacks are outlined. The analysis shows that the security of firmware updates is crucial, but has not received much attention in the literature. Therefore, the second part presents a novel firmware modification attack against one specific fitness tracker.

The first part consists of Sections~\ref{introduction} to~\ref{selected_attacks}. In Section~\ref{related_work}, related work is reviewed. Section~\ref{architecture_types} will go into more detail about how a fitness tracker works, which technologies are typically used and how the various devices work together, resulting in a general model of fitness trackers. Section~\ref{adversary_model} characterises the adversary used for the remainder of this paper. Previously published attacks on fitness trackers are surveyed in Section~\ref{selected_attacks}. The main contribution of this work -- a proof-of-concept firmware attack against Withings' collection of \emph{Activité} trackers -- is presented in Section~\ref{security_of_firmware_updates}, followed by a conclusion in Section~\ref{conclusion}.

\section{Related Work}
\label{related_work}

Related work can be divided into two important categories: Firstly, works in the narrow context of fitness trackers focussing on attacking data integrity, as well as discussions about the privacy and cultural issues surrounding their ever expanding presence are surveyed.
After that, selected works regarding firmware security and firmware modification attacks are summarised.

Kawamoto and Tsubouchi demonstrate that data from accelerometers, as also used in fitness trackers, can be used to answer highly personal questions about a user \cite{Ka14, tsubouchi2013working}, identifying the need for adequate privacy protections for users.
Paul et al.~discuss the privacy policies of various vendors, identifying with whom companies intend to share the data produced and what rights a user retains over their own data. They identify a number of worrying statements regarding the ownership and commercial usage of user data \cite{Pa14}. In 2013, Rahman et al.~examined Fitbit's communication protocol and found several weaknesses, e.g. unencrypted and unauthenticated data upload, allowing for easy data manipulation. They devised \emph{FitLock}, a ``defense system'' for the vulnerabilities discussed \cite{Rahman13}. Zhou et al.~followed up on this work by identifying shortcomings in \emph{FitLock}, but did not propose their own set of modifications to fix the aforementioned issues \cite{Zhou14}. Reexamining their work, in 2014, Rahman et al.~published a paper detailing weaknesses in both Fitbit's and Garmin's communication protocol, enabling them to send fake data to the server and inject fake data into the fitness tracker. They devised \emph{SensCrypt}, a ``secure protocol for managing low power fitness tracker'' \cite{Rahman14}. Symantec published a white paper in 2014, highlighting issues regarding the handling of private information by fitness trackers. They were able to track individuals using cheap, off-the-shelf hardware by simply connecting to fitness trackers on the street.

Costin et al.~in 2014 performed a large scale analysis of the security of embedded firmware. They presented a correlation technique that allows for propagation of existing vulnerability information to similar firmware that has previously not been identified as vulnerable. One downside of their approach is that it is rather granular and cannot identify previously unknown firmware modification attacks \cite{Costin14}. Cui et al.~present a general discussion of firmware modification attacks and then focus their work on HP's LaserJet printers, identifying an issue allowing an adversary with print permissions to update the affected printer using modified firmware \cite{CuiCS13}. Coppola in 2013 studied the firmware update procedure of Withings' \emph{WS-30} wireless scale. He combined hardware and software reverse engineering to exploit a critical flaw in the update process, allowing him to upload arbitrary firmware to the scale \cite{Coppola}.

Coppola's unpublished work on Withings' \emph{WS-30} is closest to our work in terms of content. Because fitness trackers process much more sensitive information, use a different communication architecture compared to Withings' scales, and it has been more than 2 years since Coppola demonstrated his work, firmware security on Withings' platform should have improved, warranting an analysis of the security of firmware updates in the context of fitness trackers. Such an analysis in this context has so far been overlooked in academia and, to the best of our knowledge, has not been published before.

\section{Architecture Types}
\label{architecture_types}

Fitness trackers are small, light-weight pieces of technology typically worn on the wrist. There are however also trackers that can be worn on the shoe, the waist, or the upper arm.

Priced anywhere from \$15 (Xiaomi MiBand) to over \$150 (Fitbit Charge HR), all trackers have the same core function: Counting steps. In fact, as can be seen from Tab.~\ref{tab:fitness_comparison}, the only sensor found in every fitness tracker listed is an accelerometer. This sensor is used to infer a number of data points during the day such as number of steps taken, calories burned, distance travelled, as well as time slept during the night. High-end trackers such as Fitbit's \emph{Charge HR} typically feature a small display that shows time, daily statistics and notifications from a connected phone. Additionally, this device also includes an altimeter to measure the number of flights of stairs and an optical heart rate sensor to continuously measure the wearer's heart rate \cite{Fitbit15}. Many trackers also feature a vibration motor to silently wake up the wearer.

\begin{table}
\caption{Various fitness trackers along with the sensors built in.}
\label{tab:fitness_comparison}
\centering
\begin{tabular}{c|c|c|c|c|c}
	\ & Display & Accelerometer & Heartrate & Altimeter & Vibration motor  \\\hline
	Fitbit Charge HR & \checkmark & \checkmark & \checkmark & \checkmark & \checkmark \\
	Misfit Flash & \xmark & \checkmark & \xmark & \xmark & \xmark \\
	Xiaomi MiBand & \xmark & \checkmark & \xmark & \xmark & \checkmark \\
	Withings Activité & \xmark & \checkmark & \xmark & \xmark & \checkmark
\end{tabular}
\end{table}

Due to power and space constraints, all vendors have chosen to use Bluetooth Low Energy (BLE) instead of WiFi as their communication channel. After pairing the tracker with a suitable smartphone, tablet or computer (hereafter referred to as ``Personal Communcations Device'' or simply PCD), data is transferred between tracker and an app on the PCD via Bluetooth Low Energy. For more details regarding Bluetooth Low Energy, refer to \cite{BLE1} and follow-up works where the general architecture and communication patterns are summarised.

\begin{figure}[htb]
	\caption{Communication architectures of fitness trackers}
	\label{fig:architectures}
	\centering
	\includegraphics[width=.9\textwidth]{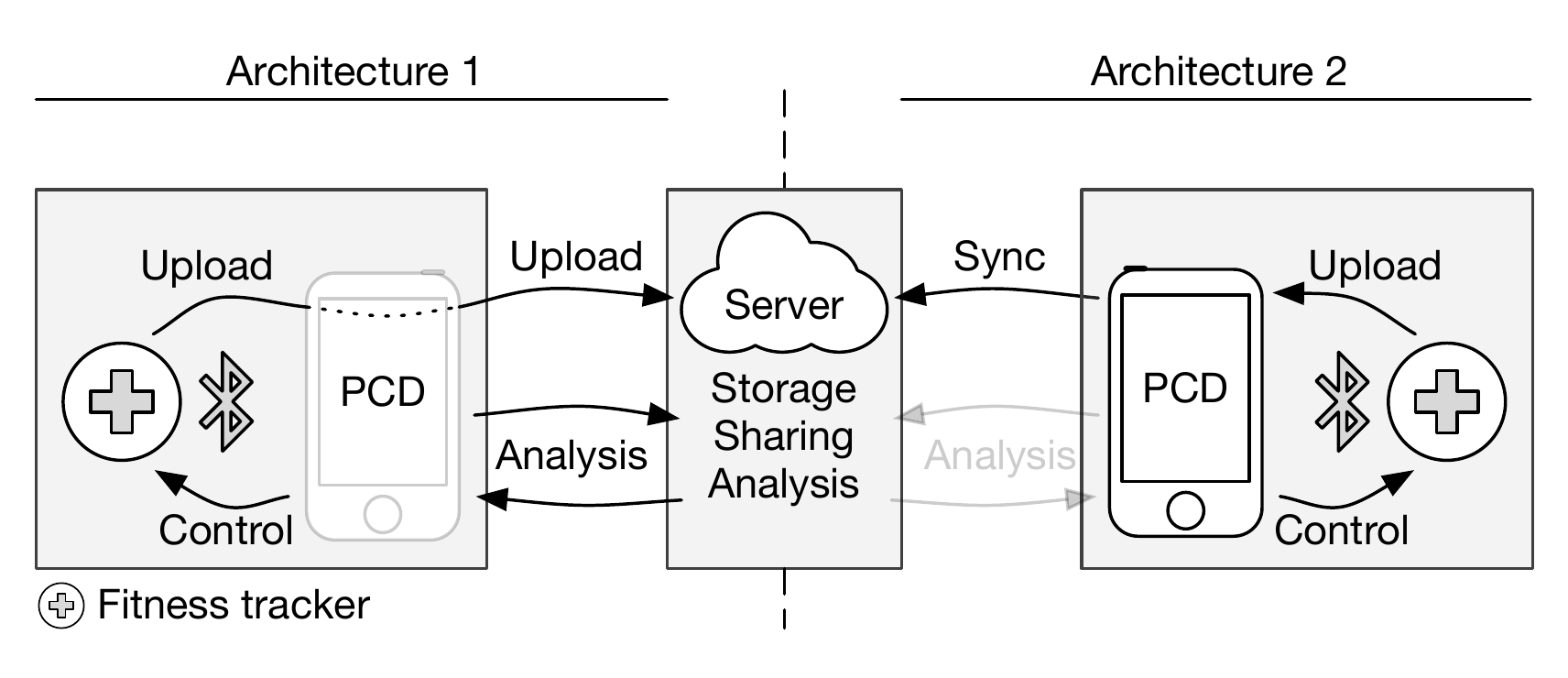}
\end{figure}

Broadly speaking, as shown in Fig.~\ref{fig:architectures}, there are two different viable communication architectures. 

The left-hand side depicts an architecture where the software running on the PCD is not capable of interpreting the data received and acts as a ``dumb'' pipe between tracker and backend server. The PCD in turn has to query the server for ``insights'' and visualisations of the data. Thus, in this architecture, the server serves a very important role. Most notably, \emph{Architecture 1} enables tracker and server to communicate end-to-end encrypted. In contrast, the right-hand side visualises an architecture where the software on the PCD is capable of interpreting data sent by the tracker, enabling visualising the data without calling on the server. In such an architecture, the server is still used for syncing, sharing and more advanced analysis.

Both architecture types -- or slight variations of them -- are used in practice: Fitbit, following a string of attacks abusing unencrypted data communications between PCD and server, now seems to be using an approach much closer \emph{Architecture 1}, where the applications itself only passes on the data \cite{FitbitReimplemented, cyr2014security}. In contrast, based on the fact that Withings' \emph{Health Mate} application can show detailed statistics even without a working Internet connection, this platform is much more similar to the second architecture.

\section{Adversary Characterisation}
\label{adversary_model}

This section presents a concise, informal description of the adversaries' motives and capabilities. Specifying a formal \emph{attacker model} makes little sense in this context, as fitness trackers themselves are not rigorously formalised.

From the point of view of the tracker's vendor -- which is considered to be a benign actor -- the tracker's \emph{user} is the primary adversary. Exercising control over both tracker and PCD and being motivated by possible financial gain, this adversary has both means and motives to try to attack the system.

The user is considered to have full control over their PCD -- including the ability to execute arbitrary code (and therefore being able to tamper with the vendor's code). This control allows for passive sniffing attacks on both wireless channels -- Bluetooth and WiFi -- to eavesdrop on communications between server, PCD, and tracker. Active attacks, such as manipulations of traffic by injecting or modifying data or denying service altogether are possible as well. Indeed, even TLS encrypted communications can be intercepted and manipulated by using a web debugging proxy (see also Sect.~\ref{selected_attacks}).

The adversary is thought to perform only software-based attacks that do leave visible traces of tampering on the target tracker itself -- Hardware based attacks, such as physically dumping firmware or data, are often very challenging, as microcontrollers use internal memory and ship with code protection features. Hardware reverse engineering is therefore explicitly \emph{not} considered. By default, the adversary is only able to use the tracker as intended, and tries to abuse his control over the other parts of the system to manipulate data or to gain full control of the tracker itself.

Other adversary models were discussed in the past: A malicious vendor would trivially be able to create malicious firmware or could modify data at will. The analysis of third-party adversaries manipulating communication channels (e.g. WiFi and / or Bluetooth) has been part of previous work (see Sect.~\ref{related_work}) and would require a thorough discussion of the communication medium itself, which is beyond the scope of this work.

\section{Attacks on Data Integrity}
\label{selected_attacks}
This section surveys existing attacks on data integrity. So far, attacks have mainly attacked exposed air channels: Both the connection between PCD and server and the local connection between PCD and tracker can be targeted. This section investigates these issues in more detail.

The first point of attack, Internet traffic, was used by practically all previous work on this matter: In their 2013 paper \cite{Rahman13}, Rahman et al.\ investigate the state of security for the Fitbit \emph{One}. They found that all traffic between basestation and the vendor's website was unencrypted, allowing them to upload modified data to the Fitbit website. Following up on their work, they identified a similar vulnerability in Garmin's web service \cite{Rahman14}. Using TLS should fix this problem, however, for the adversary as outlined in Section~\ref{adversary_model}, even a correctly implemented TLS application does not present a significant hurdle. Generally, the workflow to inspect and modify TLS traffic on one's own devices goes as follows: A web debugging proxy with TLS support, e.g. \emph{Fiddler} \cite{Fiddler} or \emph{mitmproxy} \cite{mitmproxy}, is installed. A custom root certificate is installed on the user's PCD allowing the proxy to generate TLS certificates on the fly for each requested website. Because the device in question trusts the custom root certificate, it also trusts all generated TLS certificates. This is a side-effect of the way the trust model works for TLS and is \emph{not} strictly speaking a vulnerability. If employed, \emph{certificate pinning} can be bypassed by patching the binaries, as demonstrated in practice \cite{Kramer15, Rizzo12}. For the considered adversary, TLS therefore does not pose a significant challenge.

For the Bluetooth link, similar attacks are possible. Because the considered adversary is one of the communicating parties, she does not need to break any existing protocols. In some instances, even an adversary that does not control the user's PCD can attack or at least monitor the Bluetooth traffic of fitness trackers: As detailed in \cite{Rahman14}, missing authentication allows one to harvest information of bystanders or to write bogus data to the tracker itself. Reports indicate that such attacks are still possible today \cite{Symantic}.

The last attack vector is of physical nature, where the tracker is manipulated in a way that recorded activities are not actually performed by the legitimate owner. Such ``mule-attacks'' \cite{Rahman13} make the tracker record a correct measurement of fabricated facts, violating the security property \emph{veracity} as proposed by Gollmann \cite{gollmann2012veracity}. This type of attack is not specific to fitness trackers: In much the same way, temperature readings can be manipulated and smoke alarms set off by temporarily manipulating the sensor's local environment. \emph{Unfit-Bits} (\url{http://www.unfitbits.com/}) is one parody website solely devoted to detailing ways of fooling fitness trackers \cite{Unfitbits}.

\section{Security of Firmware Updates: A case-study of Withings' \emph{Activité}}
\label{security_of_firmware_updates}

So far attacks discussed in this paper have attacked two aspects: The first kind of attack interferes with WiFi and Bluetooth channels to manipulate readings after they have been generated on the tracker. Secondly, ``mule-attacks'' manipulate the local environment to fool sensors to record steps and activities that are not performed by the legitimate owner. Both attacks allow a determined adversary to achieve the goal of data manipulation. However, both techniques require the adversary to actively manipulate the tracker over the desired period of time. Vendors have already started addressing problems abused by attacks discussed previously. Fitbit for example seems to have completely re-designed their communication infrastructure to prevent attacks abusing exposed air channels \cite{FitbitReimplemented}. We expect that vendors will also try to detect and prevent mule attacks in the future.

In contrast, attacks on firmware -- software running directly on the microprocessor embedded in these devices -- are much more capable. Such attacks can also be used to manipulate the step count and activity levels, simply by outputting incorrect information in the first place. Unlike other attacks, a firmware attack requires larger up-front costs, but scales much better: After the initial injection, no more unusual interactions with the tracker are required. In this sense, they are more convenient for an adversary. Furthermore, every single aspect of the device that is accessible by firmware can also be controlled by the adversary, including Bluetooth functionality and data storage. Firmware attacks are interesting to other adversaries apart from the user, as they offer features such as stealthiness and persistence that none of the other attacks offer.

In the following, a proof-of-concept firmware attack against Withings' \emph{Activité} fitness tracker is presented.

\subsection{Methodology}
\label{methodology}

The most straight-forward method to inject custom firmware is to abuse official update procedures. In order to exploit a flaw in this process, two basic requirements have to be met. First off, the firmware has to be transferred to the device over a (wireless) channel under the control of the adversary. This holds true, because the user controls the PCD interacting with the tracker via Bluetooth. Secondly, there have to be flaws in the verification of firmware updates. Because the PCD is under control of the adversary, effective verification and authentication can only be performed on the tracker itself.

To develop the proof-of-concept attack presented here, a number of challenges had to be overcome: The (unencrypted) firmware had to be dumped and analysed. A new image had to be crafted and re-signed to pass verification on the tracker. Finally, the modified firmware update had to be injected, preferably using the official update procedure.

To distribute updates, Withings uses the \emph{Health Mate} application on iOS and Android. All analysis presented in this paper was conducted using \emph{Health Mate} version 2.6.2 on Apple's iOS. When an update for the tracker is available, the user is notified on the main screen of the application. The app automatically downloads new firmware in the background, so that when the user initiates the update, the firmware file is transmitted wirelessly over Bluetooth onto the tracker, where it is subsequently installed. The downloaded file is not encrypted, which was immediately obvious because debug strings are readily available.
To view and modify traffic between PCD and server, \emph{mitmproxy}, a free web debugging proxy with TLS support, was used \cite{mitmproxy}.

The firmware analysis consisted of disassembling the firmware in \emph{IDA Pro} \cite{IDAPro}. To use \emph{IDA} effectively, some information such as the processor architecture used and the load address of the input file in memory are required. 

Information regarding the processor can usually be found by examining the user's manual, publicly available FCC documents or public teardowns. Indeed, one public teardown \cite{WithingsTeardown} identifies a variant of the \emph{nRF51822} as being used. This is a low-power SoC (System on a Chip) combining Bluetooth Low Energy with a 32-bit ARM Cortex-M0 which is capable of executing code that uses the \emph{ARMv6-M} instruction set \cite{nRF51822, ARMCortexM0}.

The second step, identifying a firmware's load address in memory, is much more challenging. Fortunately, previous work by Coppola addresses this problem \cite{Coppola}. A list of candidate load addresses can be built up by iteratively trying different base addresses and, judging by the number of strings referenced in the disassembly using this base address, choosing the best match.
Because two distinct binaries were packed into one file and both files were loaded at different addresses in memory, the aforementioned approach only yielded the correct load address for the first, larger image. Therefore, as a third step, the second image's load address was determined by identifying almost identical debug functions in both binaries that referenced identical strings. An approach to finding base addresses automatically makes use of jump tables and indirect jumps and is described in detail in \cite{shoshitaishvili2015firmalice}.

Manual reverse engineering is a tedious, error-prone task. With no symbol information available, as is the case for the task at hand, the analyst is left with a difficult job. Fortunately, large parts of Withings' firmware contain debug strings, making it much easier to identify code sections of interest.

\subsection{Results}
\label{results}

By reverse engineering the code within firmware images, it was possible to identify a header structure at the beginning of each firmware update that is used for firmware verification.  Figure~\ref{fig:header_structure} shows parts of the reconstructed header structures. After the header's version number -- which so far has always been \texttt{1} for all firmware updates examined -- and it's length -- which does not include the final \texttt{table\_checksum} field -- are two structures describing each of the two images that are bundled within an update, including their respective offset, length and version number. There's also a checksum field that is validated by computing a checksum over the actual contents of the image, not the header itself. Finally, the so-called \texttt{table\_checksum} is computed over the contents of the header. Based on debug strings in the update, the two binaries are internally referred to as ``app'' and ``bootloader''. Together, these binaries make up the firmware of the \emph{Activité}.

\begin{figure}[htb]
\caption{Header structure definition}
\label{fig:header_structure}
\begin{minipage}{.5\textwidth}
	\begin{lstlisting}[language=C, frame=r]
struct activite_header_t {
    uint16_t table_ver;
    uint16_t table_len;
    image_t  images[2];
    uint32_t table_checksum;
};
	\end{lstlisting}
\end{minipage}
\begin{minipage}{.5\textwidth}
	\begin{lstlisting}[language=C, frame=]
  struct image_t {
      uint16_t identifier;
      // ...
      uint32_t checksum;
      uint32_t version;
  };
	\end{lstlisting}
\end{minipage}	
\end{figure}

Checksums -- \emph{CRC-32} is used here -- can safeguard against transmission errors, but they cannot be used to verify the \emph{authenticity} of an update.
An adversary can just replace the checksums after modifying the original firmware to circumvent all checks on the tracker itself. There is nothing inside the update file that could be used to verify the authenticity of an update. This reveals a major design flaw in Withings' firmware security, which relies solely on the paradigm of \emph{Security through Obscurity}.

Crafting a modified firmware image is not much of a challenge. Any modification can be used to demonstrate the issue. Modifications such as manipulating the step count are comparatively simple, whereas complex, malicious modifications are harder and require substantial (re-)engineering resources. At this point, all modifications have to be performed by patching the firmware file directly. In the future, it might be possible to adopt work by Wang et al.~to introduce more complex modifications by recompiling the disassembly, rather than patching individual bytes \cite{ReassembleableDisassembling}.

As a proof-of-concept, the version number inside the update was modified, the firmware ``resigned'' and injected (see next paragraph for details). The tracker successfully accepted the update and transmitted the modified version number to the \emph{Health Mate} application.

\subsection{Impact}
\label{impact}
The described attacked can be performed by any adversary capable of manipulating HTTP traffic -- provided an official update is available.

\begin{figure}[htb]
	\caption{Attacking update procedure}
	\label{fig:attacking_update_procedure}
	\centering
	\includegraphics[width=.70\textwidth]{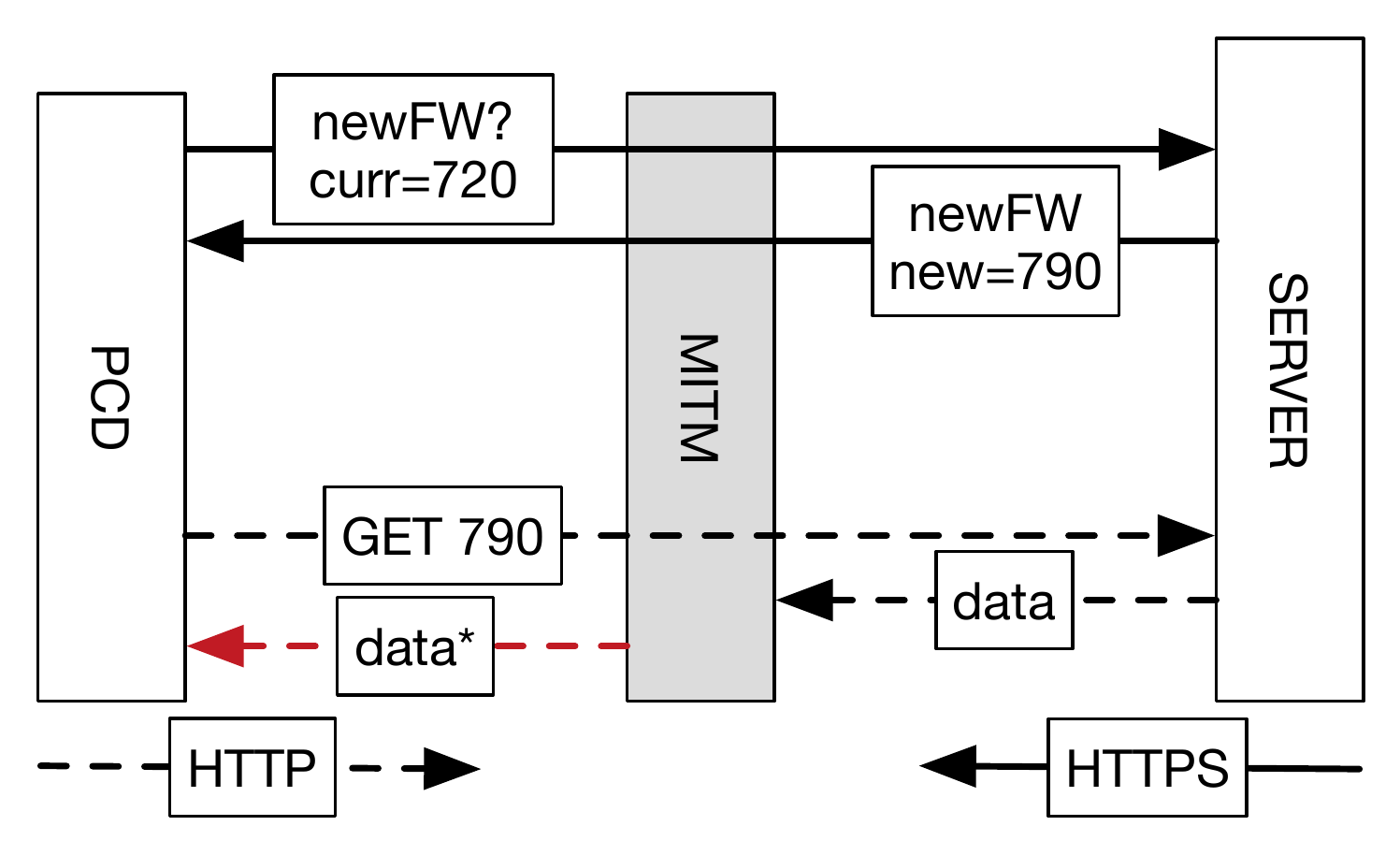}
\end{figure}

On each start of the \emph{Health Mate} application, a \emph{GET} request is made to Withings' server to check for the availability of new firmware. The response includes a URL to new firmware, if one is available. Figure~\ref{fig:attacking_update_procedure} depicts the update process along with the actions of a third-party capable of manipulating plain HTTP traffic. What can be seen is that such an adversary is still able to serve malicious updates, provided an official update is available. The considered adversary (see Sect.~\ref{adversary_model}) is more powerful: She can additionally fake the availability of new firmware using a TLS proxy and thus does not need to wait for official updates to inject custom firmware.

The discovered flaw and resulting attack apply to both Withings' \emph{Activité} and \emph{Activité Pop} and presumably \emph{Activité Steel} (At the time of writing, no device for testing was available). Analysis of the firmware of Withings' \emph{Pulse Ox} reveals the same underlying design defect. Coppola showed in 2013 that Withings' \emph{WS-30} wireless scale also did not verify the authenticity of firmware updates \cite{Coppola}. Because ARM's Cortex-M0 cores lack any kind of privilege separation \cite[p.33]{ARMCortexM0}, malicious updates have control over the whole chip and can block further legitimate updates or dynamically patch new firmware updates for persistence.

The results are potentially also transferrable to trackers of other vendors because of the similarities in the design of fitness trackers -- Misfit's \emph{Flash} for example uses the same microprocessor as Withings' \emph{Activité}.

\subsection{Countermeasures}
\label{countermeasures}

The main problem identified in the previous section is the lack of authenticity checks in Withings' firmware updates.

There are two routes to improve the current situation: As a first step, it would be wise for Withings to remove all debug strings from production firmware. While this does not fix the underlying issue, it makes it a lot harder to reverse engineer firmware. Furthermore, encrypting the firmware and decrypting the firmware on the tracker itself -- \emph{not} on the PCD which is controlled by an adversary -- would mean an adversary has to physically attack the tracker to obtain the decrypted firmware.

Fundamentally though, the authenticity of firmware has to be verified by the tracker itself. Two options come to mind: Either a unique, symmetric key is shipped with every device which then can be used to verify the firmware's \emph{MAC} (message authentication code), or a public key is embedded on the device or in the firmware itself, which is then used to authenticate subsequent updates. The symmetric key required to verify a \emph{MAC} has to be unique, otherwise a single compromised device jeopardises the security of all devices. Because each device would then have to receive individual firmware updates, shipping a public key inside the firmware might be the more cost-effective alternative. Public key cryptography however poses a significant challenge for slow, battery-constrained devices. These challenges are beyond the scope of this paper.

Finally, it is important to mention that properly authenticated firmware updates, as important as they are, do not make firmware ``secure''. Vulnerabilities in the code itself can still be exploited to attack firmware.

Countermeasures proposed here are solely meant to improve the security of firmware updates. Mitigations implemented on the PCD (e.g. certificate pinning) can only help defend against adversaries who do not control the PCD and are therefore not mentioned here.

\section{Conclusion}
\label{conclusion}

Being able to trust the authenticity and integrity of fitness data generated by trackers is essential for a number of use cases such as presenting the data as evidence in court and adjusting premiums for insurants. A survey on existing attacks on data integrity for trackers reveals that thus far, firmware level attacks were overlooked by the academic community.

In this paper, a vulnerability in the firmware update process for Withings' \emph{Activité} was identified that can be used to inject malicious code into the tracker. This is all the more worrying considering that due to limitations imposed by hardware design, there is no privilege separation on this particular tracker family.

As of now, modifications to Withings' firmware are fairly limited in scope and impact, due to the fact that manual binary patching has to be used. In the future, work by Wang et al.~might allow for much easier development of custom firmware, by recompiling disassembled firmware. Reverse engineering and documenting more functions in official updates -- not just parts that are used for verifying updates -- could allow for much more extensive modifications.

Most importantly, further research investigating other fitness trackers with respect to firmware security is needed.

\bibliographystyle{lnig}
\bibliography{fitness_paper}

\end{document}